\documentclass[useAMS,usenatbib]{mn2e}
\usepackage{graphicx}
\usepackage{graphics}
\usepackage{rotating}
\usepackage{amsmath}
\usepackage{amssymb}
\usepackage{color}
\usepackage{caption}

\newcommand{\tgss}{TGSSADR~J183304.4$-$384046}
\newcommand{\mJybeam}{\ensuremath{{\rm mJy\,beam}^{-1}}}

\def\USydney{$^{1}$}
\def\CAASTRO{$^{2}$}
\def\UWisc{$^{3}$}
\def\CASS{$^{6}$}
\def\Curtin{$^{7}$}
\def\Melb{$^{8}$}
\def\UWA{$^{17}$}

\title[Search for low-frequency radio transients]{A search for long-timescale, low-frequency radio transients}
\author[Murphy et al. ]{\parbox[t]{\textwidth}
{Tara Murphy\USydney$^,$\CAASTRO\thanks{E-mail: tara.murphy@sydney.edu.au},
David L. Kaplan\UWisc, 
Steve Croft$^{4,5}$, 
Christene Lynch\USydney$^,$\CAASTRO,
J. R. Callingham\USydney$^,$\CASS$^,$\CAASTRO,
Keith Bannister\CASS, 
Martin E. Bell\CASS$^,$\CAASTRO, 
Natasha Hurley-Walker\Curtin, 
Paul Hancock\Curtin$^,$\CAASTRO,
Jack Line\Melb,
Antonia Rowlinson$^{9,10}$,
Emil Lenc\USydney$^,$\CAASTRO,
H. T. Intema$^{11,12}$,
P. Jagannathan$^{11,13}$,
Ronald D. Ekers\CASS$^,$\Curtin,
Steven Tingay\Curtin$^{,14}$,
Fang Yuan$^{15,2}$,
Christian Wolf$^{15,2}$,
Christopher A. Onken$^{15,2}$,
K.~S.~Dwarakanath$^{16}$,
B.-Q.~For\UWA,
B.~M.~Gaensler$^{1,18,2}$,
L.~Hindson$^{19}$,
M.~Johnston-Hollitt$^{19}$,
A.~D.~Kapi\'{n}ska\UWA$^,$\CAASTRO,
B.~McKinley\Melb$^,$\CAASTRO,
J.~Morgan\CAASTRO,
A.~R.~Offringa$^{9}$,
P.~Procopio\Melb$^,$\CAASTRO,
L.~Staveley-Smith\UWA$^,$\CAASTRO,
R.~Wayth\Curtin,
C.~Wu\Curtin,
Q.~Zheng$^{19}$
}\\
\vspace*{1pt} \\
$^{1}$Sydney Institute for Astronomy, School of Physics, The University of Sydney, NSW 2006, Australia\\
$^{2}$ARC Centre of Excellence for All-sky Astrophysics (CAASTRO)\\
$^{3}$Department of Physics, University of Wisconsin--Milwaukee, Milwaukee, WI 53201, USA\\
$^{4}$Astronomy Department, University of California, Berkeley, 501 Campbell Hall \#3411, Berkeley, CA 94720, USA\\
$^{5}$Eureka Scientific, Inc., 2452 Delmer Street Suite 100, Oakland, CA 94602, USA\\
$^{6}$CSIRO Astronomy and Space Science (CASS), Marsfield, NSW 2122, Australia\\
$^{7}$International Centre for Radio Astronomy Research, Curtin University, Bentley, WA 6102, Australia\\
$^{8}$School of Physics, The University of Melbourne, Parkville, VIC 3010, Australia \\
$^{9}$Netherlands Institute for Radio Astronomy (ASTRON), PO Box 2, 7990 AA Dwingeloo, The Netherlands\\
$^{10}$Anton Pannekoek Institute, University of Amsterdam, Postbus 94249, 1090 GE, Amsterdam, The Netherlands\\
$^{11}$National Radio Astronomy Observatory, 1003 Lopezville Road, Socorro, NM 87801-0387, USA\\
$^{12}$Leiden Observatory, Leiden University, Niels Bohrweg 2, NL-2333CA, Leiden, The Netherlands\\
$^{13}$Department of Astronomy, University of Cape Town, Private Bag X3, Rondebosch 7701, Republic of South Africa\\
$^{14}$INAF, Istituto di Radio Astronomia, Via Piero Gobetti, 41029, Bologna \\
$^{15}$ Research School of Astronomy and Astrophysics, Australian National University, Canberra, ACT 2611, Australia.\\
$^{16}$Raman Research Institute, Bangalore 560080, India\\
$^{17}$International Centre for Radio Astronomy Research (ICRAR), University of Western Australia, Crawley, WA 6009, Australia \\
$^{18}$Dunlap Institute for Astronomy \& Astrophysics, University of Toronto, 50 St George St, Toronto, ON, M5S 3H4, Canada\\
$^{19}$School of Chemical \& Physical Sciences, Victoria University of Wellington, Wellington 6140, New Zealand}

\begin{document}

\date{April 2016}

\pagerange{\pageref{firstpage}--\pageref{lastpage}} \pubyear{2016}

\maketitle

\label{firstpage}

\begin{abstract}
We present a search for transient and highly variable sources at low radio frequencies (150--200~MHz) that 
explores long timescales of 1--3 years. We conducted this search by comparing the TIFR GMRT Sky Survey Alternative
Data Release 1 (TGSS~ADR1)
and the GaLactic and Extragalactic All-sky Murchison Widefield Array (GLEAM) survey catalogues.
To account for the different completeness thresholds in the individual surveys, we searched
for compact GLEAM sources above a flux density limit of 100~mJy that were not present 
in the TGSS~ADR1; and also for compact TGSS~ADR1 sources above a flux density limit of 200~mJy
that had no counterpart in GLEAM. 
From a total sample of 234\,333 GLEAM sources and 275\,612 TGSS~ADR1 sources
in the overlap region between the two surveys, there were 99\,658 GLEAM sources and 38\,978 TGSS~ADR sources
that passed our flux density cutoff and compactness criteria. Analysis of these sources resulted
in three candidate transient sources.
Further analysis ruled out two candidates as imaging artefacts. 
We analyse the third candidate and show it is likely to be real, with a flux density of $182\pm26$~mJy at 147.5~MHz. 
This gives a transient surface density of $\rho = (6.2\pm6)\times10^{-5}$~deg$^{-2}$.
We present initial follow-up observations and discuss possible 
causes for this candidate. 
The small number of spurious sources from this search demonstrates the
high reliability of these two new low-frequency radio catalogues.
\end{abstract}

\begin{keywords}
radio continuum: general --- catalogues --- galaxies: active
\end{keywords}

\section{Introduction}
There are a range of astronomical phenomena that are known to be transient
or highly variable at low frequencies ($<1$~GHz).
For example, flares from brown
dwarf stars \citep[e.g.][]{berger06,jaeger11}, flares from Jupiter \citep{zarka01}, and by extension, 
potentially from exoplanet emission \citep{hess11,murphy15}, and
intermittent pulsars \citep[e.g.][]{sobey15}. In other cases, propagation effects such as
interplanetary scintillation \citep[e.g.][]{kaplan15} can cause compact 
background sources
such as quasars and pulsars to vary in flux density. There are also more local 
causes such as ionospheric distortions. Although these are typically
small effects at a few hundred megahertz at the resolution of the Murchison
Widefield Array \citep{loi15b} they can be significant for higher resolution 
instruments \citep{intema09,vanweeren16}. The range of physical phenomena that cause 
radio variability is summarised by \citet{cordes04} and by \citet{bowman13} for low 
frequencies in particular.

Most studies of low frequency variability have targeted known objects, however
there have been a small number of transients discovered through limited blind searches \citep[e.g.][]{hyman05}.
In recent years large-scale blind transient surveys have become possible due to
new instruments such as the Long Wavelength Array \citep[LWA1;][]{taylor12},
the Low Frequency Array \citep[LOFAR;][]{vanhaarlem13}, the Murchison Widefield
Array \citep[MWA;][]{tingay13} and the Jansky Very Large Array (VLA) Low Band 
Ionospheric and Transient Experiment \citep[VLITE;][]{clarke16}. A range of 
low-frequency surveys for radio transients have been conducted on these 
instruments 
\citep[e.g.][]{bell14,obenberger15,polisensky16} in part as 
preparation for the Square Kilometre Array \citep[SKA; e.g.][]{fender15}. 
In addition there have been a number of 
surveys using archival low frequency data from the Very Large Array 
\citep{jaeger12} and Molonglo \citep{bannister11} telescopes.

So far, most surveys for radio transients at low frequencies \citep[e.g.][]{lazio10,bell14,polisensky16} have not found any astrophysical sources,
resulting in upper
limits on the rate of transient and highly variable sources as summarised by 
\citet{rowlinson16}. However, there are a few surveys that have resulted in detections.
One is the archival search of 325~MHz Very Large Array observations by \citet{jaeger12}.
They found a single candidate in 72~hours of observations in the Spitzer-Space-Telescope Wide-area Infrared Extragalactic Survey (SWIRE) Deep Field. This source was detected
over a 6 hour period with a flux density of $1.70\pm0.25$~mJy. 
Another example is the recent result by \citet{stewart16} who found
a single candidate astrophysical transient in 400~hr of LOFAR monitoring of the North Celestial
Pole. Their candidate had a duration of a few minutes and a 60~MHz flux density of $15-25$~Jy.
The nature of this object is still unknown, but it suggests that we might now be approaching
the survey parameters required to detect low-frequency radio transients in blind surveys.
\citet{metzger15} present model predictions for the rates of extragalactic synchrotron transients
which are significantly lower than the rates that current low-frequency blind surveys probe. However,
these do not cover all the classes of objects we expect to see at low frequencies, in particular 
Galactic objects or coherent emitters such as intermittent pulsars.

In this paper we present a search for radio transients at $\sim 150$~MHz, over 
typical timescales of three years, to a flux density cutoff of 100--200~mJy. We 
conducted the search using the GLEAM survey \citep{wayth15} survey and the 
first Alternative Data Release of the 150 MHz TIFR GMRT Sky Survey \citep[TGSS~ADR1;][]{intema16}.  Unless stated otherwise, all flux density limits are at 3$\sigma$ confidence.

\section{Data analysis}
\subsection{The GLEAM Survey}

The Murchison Widefield Array \citep{tingay13}
is a 128-tile low-frequency radio interferometer located in Western Australia.
One of the major MWA projects is GLEAM: The GaLactic and Extragalactic All-sky MWA 
survey \citep{wayth15}. GLEAM is a survey of the radio sky south of declination 
$+30^\circ$ at frequencies between 72 and 231~MHz. The observations used in this work
were carried out between 2013 June and 2014 July.
At 154~MHz the image resolution is approximately $2.5^\prime \times 2.2^\prime/\cos(\delta + 26.7^\circ)$. The 
typical sensitivity of GLEAM snapshot images ranges from 40 to 200\,\mJybeam\ between 
231~MHz and 72~MHz respectively. 
The final catalogue is $92.6\%$ complete at 200~mJy and $79.5\%$ complete at 100~mJy.

The observation and data reduction strategy for GLEAM is described in detail by \citet{hurleywalker16}.

\subsection{The TGSS ADR1 Survey}
The Giant Metrewave Radio Telescope \citep[GMRT;][]{swarup90} is an array of 30 antennas, each 
with a diameter of 45~m. The GMRT has a maximum baseline of 25~km and 
operates at frequencies between 150 and 1500 MHz. 
Between 2011 April and 2012 March the GMRT was used to carry out a 150~MHz survey 
of the sky north of declination $\delta = -55^\circ$. 

Until recently, the bulk of the data collected by the TGSS was unpublished. 
Motivated by recent improvements in low-frequency calibration and imaging, 
\citet{intema16} have re-processed the TGSS observations and released both the 
resulting Stokes I continuum images and a catalog of 620\,000 radio 
sources down to  the 7$\sigma$ level (TGSS~ADR1). The TGSS~ADR1 data release covers
the sky north of declination $\delta = -53^\circ$, excluding the most southern (lowest elevation) pointings.
The images have a median RMS of 3.5\,\mJybeam\ 
and an approximate resolution of $25^{\prime\prime} \times 25^{\prime\prime}$ or 
$25^{\prime\prime} \times 25^{\prime\prime}$/$\cos(\delta - 19)^\circ$ for declinations south of 
$\delta = +19^\circ$.

\subsection{Search strategy}
\begin{figure}
\centering
\includegraphics[width=1.05\columnwidth]{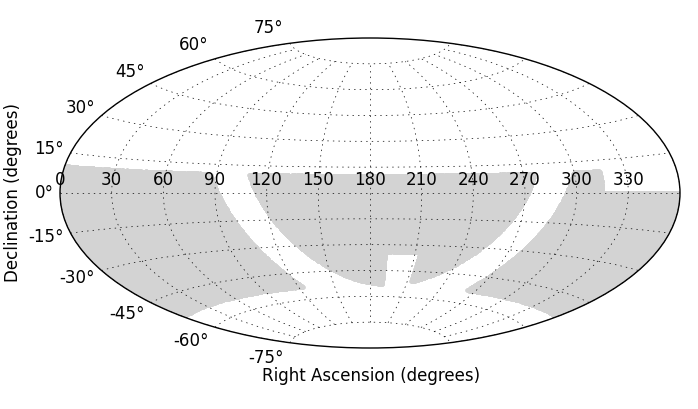}
\caption{The sky coverage of our survey (in Aitoff projection); the grey shaded area shows the overlap region between
GLEAM and TGSS, excluding the sky north of $\delta = 10^\circ$. The masked regions, described in the text,
 are shown in white.}
\label{sky}
\end{figure}

We conducted our search using the GLEAM first data release catalogue \citep{hurleywalker16} and the
TGSS ADR1 7-sigma source catalog \citep{intema16}.
The two surveys cover a common area of sky between $-53^\circ \le \delta \le +30^\circ$.
The GLEAM survey has significantly lower
completeness above $\delta = +10^\circ$ so we excluded these northern declinations from our
comparison.
The GLEAM catalogue excludes the Galactic plane region
$|b|<10^\circ$ and a 1289 sq deg area centred on $22.5^h$, $+15^\circ$ that was 
ionospherically distorted. There are also several small regions that are masked around the
bright radio galaxy Centaurus A. Figure~\ref{sky} shows the regions used 
for the analysis in this paper; the total area surveyed is 16\,230 sq deg (39.3\% of the sky).

On average the TGSS observations took place about three years before the GLEAM survey observations.
The GLEAM survey includes a sub-band centred on 154~MHz, which is close to the TGSS frequency
of 150~MHz, so they are well matched. Due to the different properties of these surveys, as discussed
below, we searched for transients using two approaches. We crossmatched sources in the GLEAM survey with 
TGSS, and looked for those with no match, and we crossmatched sources in TGSS with the GLEAM catalogue
and looked for sources with no match.

In conducting these searches there are two main differences between the survey catalogues that need to be
accounted for (i) the sensitivity limit and completeness of GLEAM compared to TGSS; and 
(ii) the lower resolution of GLEAM compared to TGSS.

(i) The typical RMS sensitivity of TGSS is 3.5\,\mJybeam, and the TGSS catalogue was constructed using 
a $7\sigma$ source-finding cutoff, which gives a limiting flux density of 24.5~mJy. The GLEAM survey 
is less sensitive; the catalogue is $92.6\%$ complete at 200~mJy, whereas TGSS is close to $100\%$ complete 
at 200~mJy.
To account for this we used a relatively high flux density cutoff of 100~mJy for
our GLEAM to TGSS comparison. For the reverse comparison we applied a flux density cutoff of 
200~mJy to the TGSS sources,
to avoid false positive transient candidates due to incompleteness in the GLEAM survey.

(ii) The resolution of the TGSS is $25^{\prime\prime}\times25^{\prime\prime}$ for $\delta>19^\circ$ and 
$25^{\prime\prime}\times25^{\prime\prime}/\cos{(\delta-19^\circ)}$ for $\delta<19^\circ$, compared 
to $2.5^\prime \times 2.2^\prime/\cos(\delta + 26.7^\circ)$ for GLEAM. The lower resolution
of GLEAM means we would expect to see a significant number of objects that have extended 
emission that is resolved out in TGSS, which produces a large number of false positives when 
crossmatching GLEAM to TGSS. A related issue is multi-component radio galaxies in which the 
catalogued position is offset between the two catalogues.

We addressed these issues firstly by restricting the crossmatch sample to sources that were
compact in GLEAM (using the criteria discussed in Section~\ref{s_g2t}); and secondly by searching for GLEAM sources that had no match in TGSS or in 
three archival catalogues: the NRAO VLA Sky Survey \citep[NVSS, at 1.4\,GHz;][]{condon98}; the Sydney University Molonglo 
Sky Survey \citep[SUMSS, at 843\,MHz;][]{mauch03}; and the VLA Low Frequency Sky Survey, redux \citep[VLSSr, at 74\,MHz;][]{lane14}.

\subsection{Flux density scale}
Before we conducted our search we checked that the flux density scales of the two catalogues
were aligned.
To do this we selected bright sources ($> 900$~mJy) in our survey region that were compact in both GLEAM 
(using the criterion $(a \times b)/(a_{\mathrm{psf}}\times b_{\mathrm{psf}}) < 1.1$ where $a$ and $b$ are the 
major and minor axis of a Gaussian fit to the source) and TGSS (using the criterion that the object was fit 
by a single Gaussian). 

\begin{figure}
\includegraphics[width=\columnwidth]{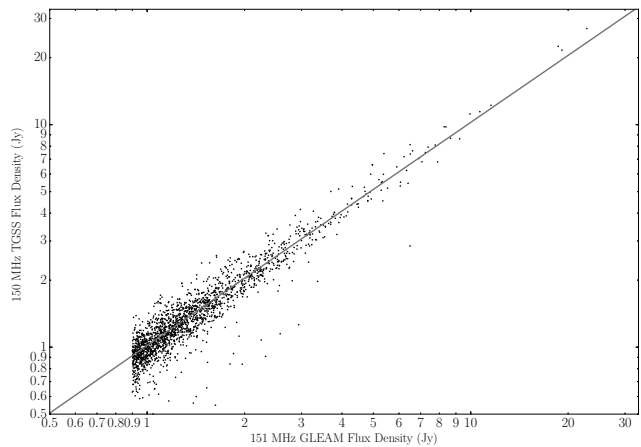}
\caption{TGSS 150~MHz flux density versus the GLEAM 151~MHz sub-band flux density for bright, compact sources
in our survey area. The line of best fit, shown as a solid line, gives a TGSS/GLEAM flux density ratio of 0.97.}
\label{fluxscale}
\end{figure}

We cross-matched these objects, and compared their flux densities as shown in Figure~\ref{fluxscale}. 
We fit these data using orthogonal distance regression, with the uncertainty in $S_{\rm TGSS}$ and $S_{\rm GLEAM}$ as weights. 
The uncertainty on $S_{\rm GLEAM}$ was the RMS noise combined in quadrature with the systematic uncertainty.
For the $\sim3000$ sources plotted, the best fit ratio of TGSS/GLEAM was 0.97, demonstrating the systematic 
difference between the measured flux densities in these catalogues is $\sim$\,$3\%$. This difference is 
to be expected since the GLEAM survey is on the \citet{baars77} flux density scale, while TGSS is on 
the \citet{scaife12} flux density scale.

Although the overall agreement is good, there is some scatter around the mean ratio, with a 
small fraction of sources showing significantly different flux densities in both surveys.
The likely cause of these differences is a combination of 
the difference in resolution between the two surveys; the better low surface brightness sensitivity of GLEAM;
errors in the local calibration and flux scale in each survey; and some variability in the source 
population. 
In particular the flux scale agreement is different in different regions of the sky: we
discuss this further in Section~\ref{tanalysis}.

\section{Results}

\subsection{GLEAM to TGSS comparison}\label{s_g2t}
There are a total of 234\,333 GLEAM catalogue sources in the overlap region 
between the two surveys. We did an initial exploration of the data by cross-matching the GLEAM catalogue with the
TGSS, SUMSS, NVSS and VLSSr catalogues, and visually inspecting all GLEAM sources that did not have a match
in any of these catalogues (a total of 2\,219 sources). A majority of these sources were either extended, or had
a match in the survey images that was below the formal catalogue limits for a given survey.

From this exploration we developed the criteria below to search for GLEAM sources that had no TGSS counterpart: 
\begin{enumerate}
\item Applied a GLEAM 200~MHz flux density cutoff of 100~mJy ({\it leaving 133\,686 sources});
\item Excluded sources that were extended in GLEAM, using the criterion $(a \times b)/(a_{psf}\times b_{psf}) \ge 1.1$ where
$a$ and $b$ are the major and minor axis of a Gaussian fit to the source ({\it leaving 99\,658 sources});
\item Cross-matched the GLEAM catalogue with the TGSS catalogue, using a search radius of $1.0^\prime$ and 
selected sources with no match ({\it leaving 1\,371 sources});
\item Cross-matched the sources with no TGSS counterpart with the SUMSS, NVSS and VLSSr catalogues, using a 
search radius of $2.5^\prime$ and selected sources with no match ({\it leaving 15 sources});
\item Visually inspected the sources and excluded a small number of sources that were GLEAM processing artefacts.
\end{enumerate}
This resulted in two candidate transients above our 100~mJy limit, GLEAM J153424$-$114947 and GLEAM J153653$-$115052,
which we discuss below.

The sources that did not have a match in the TGSS catalogue, but that did have a counterpart in NVSS or SUMSS 
(excluded in step (iv) above) were typically multiple-component radio galaxies. These were unresolved in GLEAM, 
but resolved into
multiple components in TGSS, and hence had a catalogued position or positions that were more than $1.0^\prime$ from
the GLEAM position.

The small number of GLEAM imaging artefacts that we discovered in the manual inspection process have now been 
removed from the final GLEAM catalogue.

\subsubsection{Analysis of GLEAM candidates}
GLEAM~J153424$-$114947 has a 200~MHz flux density of $204\pm10$~mJy. There is no detection in TGSS,
with a $3\sigma$ limit of 8.4~\mJybeam. There are no detections in the VLSSr ($3\sigma$ limit of 
330~\mJybeam) or NVSS ($3\sigma$ limit of 1.4~\mJybeam) surveys.

GLEAM~J153653$-$115052 has a 200~MHz flux density of $134\pm10$~mJy. There is no detection in TGSS,
with a $3\sigma$ limit of 6.2~\mJybeam. There are no detections in the VLSSr ($3\sigma$ limit of 
231~\mJybeam) or NVSS ($3\sigma$ limit of 1.3~\mJybeam) surveys.

These detections appeared to be robust (no sign of nearby bright sources or other apparent causes), however their close proximity on the sky raised concerns and so we 
inspected the individual snapshot images from which the GLEAM mosaics are constructed. 
In the snapshot images, each source only appears only in one image.  We realized from examination of these 
and at least six additional snapshot images that there was just a single source that appeared to move across 
the sky with time, shifting by approximately $0\fdg 6$ in right ascension between adjacent images taken 
10\,min apart, with minimal motion in declination.  
The source is very close to the edge of the images ($y>3880$, with an image size of 4000\,pixels) in 
the region where deconvolution is disabled.

\begin{figure*}
\centering
\includegraphics[width=17cm]{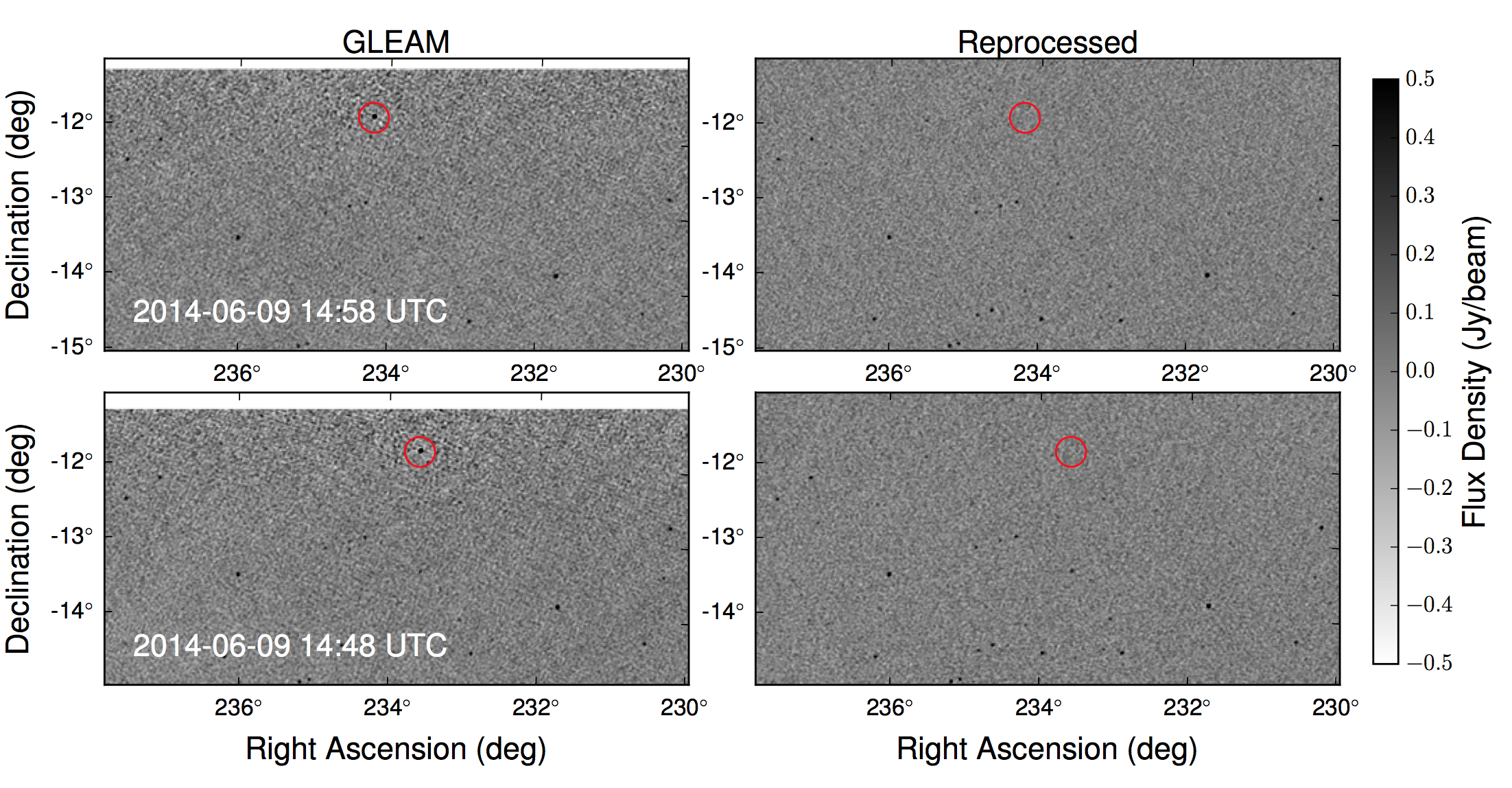}
\caption{{\it Left:} GLEAM snapshot images showing the two transient candidates GLEAM~J153424$-$114947 and GLEAM~J153653$-$115052. {\it Right:} The reprocessed images in which the candidates no longer appear, demonstrating that these are imaging artefacts.}
\label{f_artefacts1}
\end{figure*}

We reprocessed the individual snapshot images using different imaging parameters: we used the same pixel scale of 
$25\farcs2$~pixel$^{-1}$, but with an image size of 6144 pixels.
This meant that 
the imaged area was $43^\circ$ compared to $30\fdg 34$, and the source was only 85\% of the way to the edge 
rather than 97\%.  Otherwise the imaging parameters were similar.  We found no source at the position of 
GLEAM~J153424$-$114947 or GLEAM~J153653$-$115052 in either of the reprocessed snapshots where they 
previously appeared (see the right hand panels of Figure~\ref{f_artefacts1}), while the other sources 
remained, demonstrating that the potential transients are likely to be imaging artefacts. The most likely
explanation is that they are a form of aliasing in which a bright source outside the field has been aliased
in.

\subsubsection{GLEAM artefacts}
As a result of this analysis these two candidates were established as artefacts and removed from the main GLEAM catalogue.
In the GLEAM pipeline processing the primary field-of-view is imaged down to at least $10\%$ of the primary beam, leaving,
for every observation, approximately 100 square degrees of sky
with a primary beam response between $0-10\%$ outside the imaged
field-of-view. The 7 brightest sources in the sky are peeled where
possible, but in two cases, the sources could not be peeled due to
contamination from other bright sources. In one observation, Hercules A
lay just outside the field-of-view, and was not peeled, due to the
confounding presence of 3C353 in a side lobe, resulting in an alias inside
the imaged field-of-view. In 19 further cases, Centaurus A could not be
peeled due to contamination from the Galactic Plane, resulting in 19
further aliases. These were manually removed from the final catalogue. For
images which form the source-detection mosaics across 170--231\,MHz, all
other cases of potential unpeeled source contamination were checked and no
aliases were found. There are a small number of aliases in the
lower-frequency mosaics, but these are not used for source-finding, and
the aliases do not coincide with other sources.

It is possible that sources of flux densities comparable to 3C444
($S_\mathrm{200MHz}\approx60$\,Jy), the faintest peeled source, could be
just outside the field of view and cause aliases inside it. However, the
chances of this are very small, as the source must lie at a Declination
within five degrees of one of the GLEAM pointing centres, in order to fall
on an image edge rather than a corner, and fall just outside the
field-of-view. Because the GLEAM mosaics are formed by weighting snapshots
by the square of their primary beam responses, the contribution of such an
alias to the mosaic would be down-weighted by at least a factor of 100,
effectively reducing its flux density by that factor. Therefore, we do not
expect sources fainter than 3C444 to contribute detectable sources to the
mosaics, as they would appear with $S<6$\,mJy, which is below the
detection limit across the sky.

Hence, as a result of the extensive quality control done as part of the GLEAM survey, 
we do not think it is likely that similar
artefacts would have a significant impact on the overall reliability of the GLEAM catalogue. 

\subsection{TGSS to GLEAM comparison}
There were a total of 275\,612 TGSS sources in our survey region.
To do the reverse comparison, crossmatching TGSS sources with GLEAM, we followed the steps below:
\begin{enumerate}
\item Selected TGSS sources above 200~mJy ({\it leaving 74\,876 sources});
\item Selected compact sources by choosing those that were fit by a single Gaussian, marked `S' in the 
TGSS ADR1 catalogue ({\it leaving 38\,978 sources});
\item Crossmatched these with the GLEAM catalogue, using a radius of $1^\prime$ and selected sources with no
match ({\it leaving 640 sources});
\item Visually inspected all sources with no match and excluding those that were multi-component sources (generally 
double and triple radio galaxies) that are resolved in TGSS but not in GLEAM;
\item Excluded sources that are in regions of poor image quality in the TGSS mosaics.
\end{enumerate}
This resulted in a single candidate transient, TGSSADR~J183304.4$-$384046, which we discuss below.

\subsubsection{Analysis of the TGSS candidate}\label{tanalysis}
\begin{figure}
\centering
\includegraphics[width=\columnwidth]{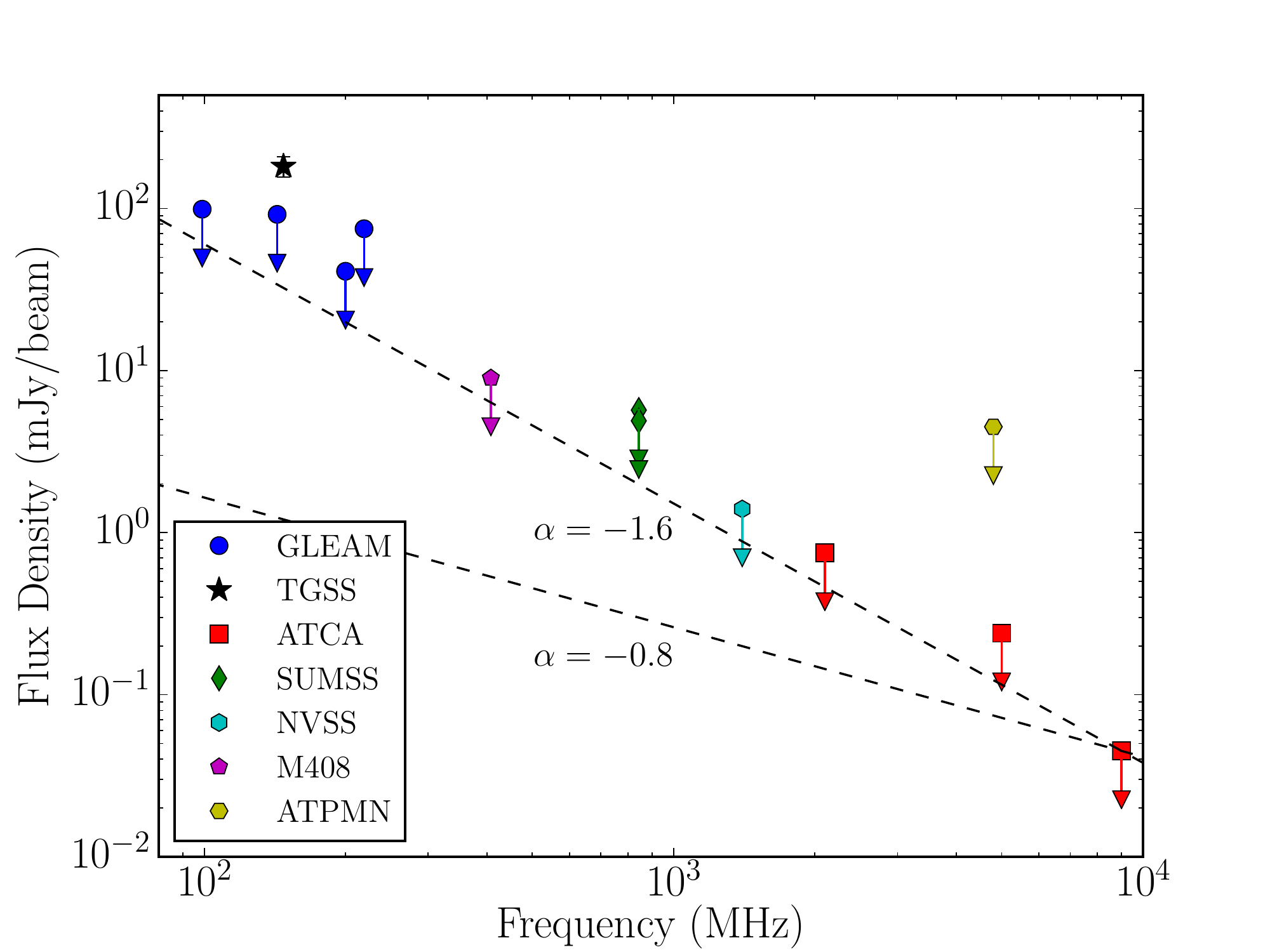}
\caption{Radio spectral energy distribution (SED) of \tgss, based on the data in Table~\ref{t_can3}. We have used the scaled TGSS flux density, as discussed in the main text. Note that these are not based on contemporaneous observations.  For the presumed quiescent emission (ignoring the TGSS detection), the most constraining measurement for typical spectral indices is the ATCA 9\,GHz observation.  We show model SEDs for spectral indices of $\alpha=-0.8$ (typical for extragalactic GLEAM sources; \citealt{hurleywalker16}) and $\alpha=-1.6$ (the steepest spectral index that matches all of the limits).
}\label{sed}
\end{figure}

\begin{figure*}
\centering
\includegraphics[width=15cm]{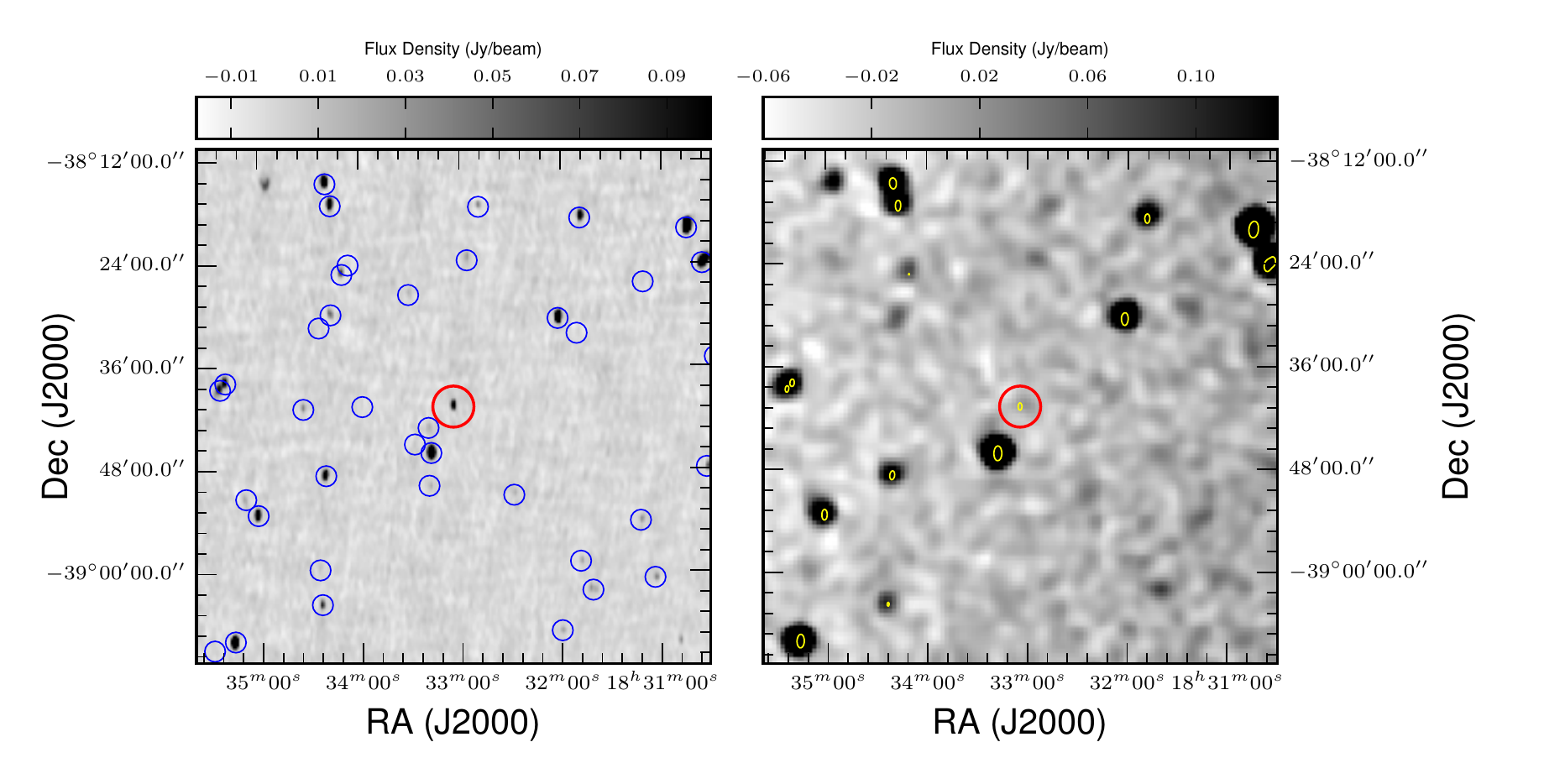}
\caption{TGSS~ADR1 (left) and GLEAM (right) images of the transient candidate TGSSADR~J183304.4$-$384046 (shown with a red circle). On the TGSS~ADR1 image, the blue circles show the location of NVSS sources with a 1.4~GHz flux density greater than 5\,mJy. The GLEAM image is overlaid with a TGSS contour at 80\,\mJybeam\ to show the position
of TGSS sources.
}
\label{f_can3}
\end{figure*}

TGSSADR~J183304.4$-$384046 has a 150~MHz flux density of 304~mJy, but was not detected in GLEAM ($3\sigma$ limit of 41~\mJybeam), SUMSS ($3\sigma$ limit of 4.9~\mJybeam), NVSS ($3\sigma$ limit
of 1.4~\mJybeam) or AT20G ($3\sigma$ limit of 30~\mJybeam). We also searched archival radio surveys ATPMN \citep{mcconnell12} and the Molonglo 408~MHz 
survey (Hunstead, private communication) and found no detections. These limits are summarised in 
Table~\ref{t_can3} and Figure~\ref{sed}. We did not detect the source
in any of the 20 GLEAM sub-bands, and we have included 4 representative sub-band limits on the flux density in the table and SED.
The TGSS and GLEAM images are shown in Figure~\ref{f_can3}. 

\begin{table}
\centering
\caption{Summary of radio measurements and limits for TGSSADR~J183304.4$-$384046.}
\label{t_can3}
\begin{tabular}{lrccl}
\hline
Survey & Freq  & Date & S$^\dagger$ &  Ref\\
       & (MHz) &      & (mJy/bm)   &  \\
\hline
GLEAM & 99    & 2014 Jun 09 & $\phantom{0}<99\phantom{.}\phantom{0}\phantom{0}\phantom{0}$ & * \\
GLEAM & 143   & 2014 Jun 09 & $\phantom{0}<92\phantom{.}\phantom{0}\phantom{0}\phantom{0}$ & * \\
TGSS  & 147.5 & 2011 Apr 27 & $182\pm26^\ddagger$ & I16 \\
GLEAM & 200   & 2014 Jun 09 & $\phantom{0}<41\phantom{.}\phantom{0}\phantom{0}\phantom{0}$ & * \\
GLEAM & 219   & 2014 Jun 09 & $\phantom{0}<75\phantom{.}\phantom{0}\phantom{0}\phantom{0}$ & * \\
M408  & 408   & 1972 Jul 15 & $<9\phantom{.}\phantom{0}\phantom{0}\phantom{0}$ & Hpc \\
SUMSS & 843   & 2005 Aug 03 & $<5.7\phantom{0}\phantom{0}$ & M03 \\
SUMSS & 843   & 2006 Mar 09 & $<4.9\phantom{0}\phantom{0}$ & M03 \\
NVSS  & 1400  & 1993 Oct 07 & $<1.4\phantom{0}\phantom{0}$ & C98 \\
ATCA  & 2100  & 2016 Apr 01 & $<0.75\phantom{0}$ & * \\
ATPMN & 4800  & 1994 Mar 14 & $<4.5\phantom{0}\phantom{0}$ & M12 \\
ATCA  & 5000  & 2016 Apr 01 & $<0.24\phantom{0}$ & * \\
ATCA  & 9000  & 2016 Apr 01 & $<0.045$ & * \\
AT20G & 20\,000 & 2004 Aug 11 & $\phantom{0}<30\phantom{.}\phantom{0}\phantom{0}\phantom{0}$ & H11 \\
\hline
\end{tabular}\\
$\dagger$ Flux density or 3$\sigma$ upper limits.\\
$\ddagger$ See Section~\ref{tanalysis} for discussion of the flux scale. \\
References: * (This work), C98 \citep{condon98}, Hpc (Hunstead, private communication), H11 \citep{hancock11}, I16 \citep{intema16}, M03 \citep{mauch03}, M12 \citep{mcconnell12}
\end{table}

Based on these non-detections we observed this source with the Australia Telescope Compact Array on 2016 April 1. We had a total
time of 1.5 hours on source at each of 2.1, 5.0 and 9.0~GHz in the 
H214 array. We did not see any emission at the position of \tgss, 
obtaining limits of 0.75~\mJybeam\ at 2.1~GHz, 0.24~\mJybeam\ at 5~GHz and 0.045~\mJybeam\ at 
9~GHz. This suggested the source was either a transient that had faded since the TGSS observations, or an artefact. 

We considered the possibility that this source was an imaging artefact from the TGSS processing, for example a CLEAN artefact
caused by the bright source approximately $6^\prime$ to the south west. This neighbouring source has a GLEAM 151 MHz flux density of $1.75\pm0.04$\,Jy (GLEAM~J183318$-$384608), and a 147.5~MHz flux density of $3.0\pm0.3$\,Jy (TGSSADR~J183317.7$-$384613) in the TGSS~ADR1.

The significant difference in the measured flux density of this neighbouring source led us to do further analysis of the flux density
scale in the region of the transient candidate. We selected all bright
compact sources within a 1~degree radius of the candidate and found the
mean flux ratio of the GLEAM 151~MHz flux density to the TGSS 147.5~MHz flux density was 0.6. This is due to flux scale uncertainties in the
TGSS~ADR that are currently being rectified. Based on this analysis, the 
flux density of the candidate transient may be $0.6\times 304 = 182\pm26$~mJy rather than the catalogued value. We have used this scaled value in the spectral
energy distribution in Figure~\ref{sed}. We have also scaled the TGSS image in Figure~\ref{f_can3} by a factor of 0.6.  We estimate that this scaling has an additional uncertainty of 10\% based on the scatter of the flux ratios.

We looked at the uncombined TGSS pointing images, before primary beam 
correction and found that the transient candidate appears in two images (R56D08 and R57D09). These pointings were observed on the same night, but with slightly different
{\it uv}-coverage. Each pointing was processed independently. Although the transient
candidate is well beyond the half-power point of the primary beam in image R57D09, it
is still clearly visible, which makes it unlikely to be an artefact. In addition, we 
did not find evidence of similar artefacts around other sources of similar brightness to this candidate's neighbour, making it unlikely that the candidate is a result of the CLEAN process. Finally, Figure~\ref{f_can3} shows that much fainter sources in the same region have matches in NVSS 
(blue circles) demonstrating the high image fidelity.
Hence after this analysis we concluded that the transient is likely to be real.

\subsubsection{Multi-wavelength Search}
\label{s_mw}
\begin{figure*}
\centering
\includegraphics[width=\textwidth]{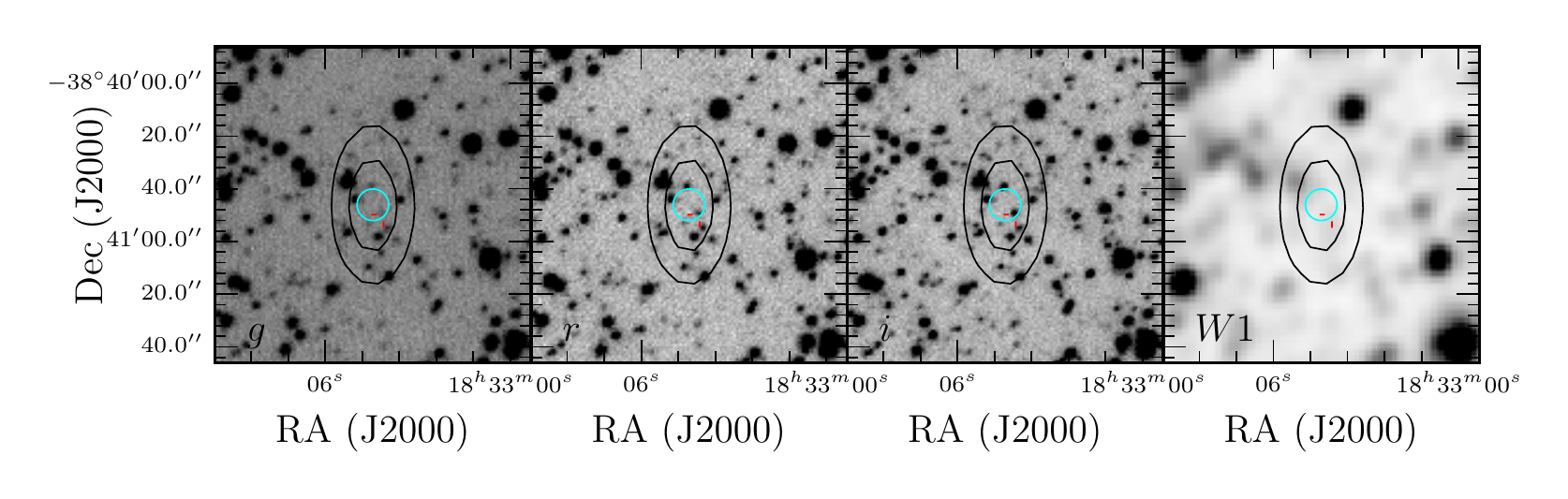}
\caption{Multi-wavelength images of \tgss, showing SkyMapper $g$, $r$, and $i$ from left to right, along with \textit{WISE} $W1$ at the right.  All images show the same field-of-view.  The contours mark 100\,\mJybeam\ and 200\,\mJybeam.  The cyan circles show the 3$\sigma$ radius ($6\arcsec$) region searched for counterparts, and the ticks mark the single stellar source identified within that region.
}
\label{f_skymapper}
\end{figure*}

We found no obvious infrared counterparts in the four \textit{Wide-field Infrared Survey Explorer} \citep[\textit{WISE};][]{wright10} bands (see Figure~\ref{f_skymapper}) or in the three 2-Micron All-Sky Survey \citep[2MASS;][]{twomass} bands, with 5\,$\sigma$ upper limits in  Table~\ref{t_oir}. We also found 
no likely counterparts in SIMBAD and NED searches. The position 
of the source rules out solar system planets.

We observed the location of \tgss\ with the 1.3\,m SkyMapper survey telescope \citep{keller07} on 2016~April~07, with $3\times 100\,$s exposures in the $g$ filter, $5\times 100\,$s exposures in the $r$ filter, and $10\times 100\,$s exposures in the $i$ filter.  The seeing was about $2\farcs5$ in all bands.  The data were processed with the standard SkyMapper pipeline and the individual images coadded.  As shown in Figure~\ref{f_skymapper}, we see only a single source within the $3\,\sigma$ radius ($6\arcsec$) error circle, down to $5\,\sigma$ limiting depths of 21\,mag in all bands.  The source appears unresolved, and has $g>21$, $r=19.80\pm0.07$, and $i=18.73\pm0.03$ (Table~\ref{t_oir}).  Using the extinction model of \citet{drimmel03} we estimate the extinction to be $A_V\approx 0.5\,$mag for distances $\gtrsim 1\,$kpc.

\begin{table*}
\centering
\caption{Summary of optical and IR measurements for TGSSADR~J183304.4$-$384046.}
\label{t_oir}
\begin{tabular}{lccccccccccccc}
\hline
& \multicolumn{3}{c}{SkyMapper$^*$} & & \multicolumn{3}{c}{2MASS$^\dagger$} & & \multicolumn{4}{c}{$WISE$$^\ddagger$}\\ \cline{2-4} \cline{6-8} \cline{10-13}
Filter & $g$ & $r$ & $i$ & &$J$ & $H$ & $K_s$ && W1 & W2 & W3 & W4\\ \hline
Wavelength ($\mu$m) & $0.48$ & $0.63$ & $0.77$ && 1.24 & 1.66 & 2.16 && 3.35 & 4.60 & 11.6 & 22.1 \\
Magnitude & $>21$ & $19.80 \pm 0.07^\S$ & $18.73 \pm 0.03^\S$ && $>16.9$ & $ >16.2$ & $>15.5$ && $ >15.7 $ & $ >15.4 $ & $ >11.6 $ & $ >8.2 $ \\
Flux Density ($\mu$Jy) & $<14$ & $44\pm3^\S$ & $117\pm3^\S$ && $<277$ & $<339$ & $<421$ && $<162$ & $<119$ & $<726$ & $<4408$ \\
\hline
\end{tabular}\\
$*$ Magnitudes are defined on the AB system.\\
$\dagger$ Magnitudes are defined on the Vega system.\\
$\ddagger$ Magnitudes are defined on the Vega system.\\
$\S$ Assuming the source discussed in the text is the counterpart. If not, the source is $r > 21$ and $i > 21$, or $F_\nu,r<15\,\mu$Jy and $F_\nu,i<15\,\mu$Jy.\\
\end{table*}

\section{Discussion}\label{s_discuss}

\subsection{Interpretation of \tgss}
There are a number of possible interpretations of this object, which we will briefly discuss here. Further analysis
is left for follow-up work once more observations have been conducted.

We have a single detection of this source, at $182\pm26$~mJy at 147.5~MHz. The non-detection in GLEAM 3 years later
implies the source has at least faded to $<41$~\mJybeam. 
We show the spectral energy distribution in Figure~\ref{sed}.  Assuming a typical spectral index of $\alpha=-0.8$ (where $S_\nu \propto \nu^\alpha$), the non-detection with the ATCA at 9\,GHz implies a 151\,MHz flux density of 1.2\,mJy, which would require considerably deeper observations to confirm.  

 The Galactic latitude of $b=-13.2^\circ$ means we must consider both Galactic and extragalactic origins for \tgss. 
We can estimate the brightness temperature $T_B$ for both scenarios: assuming variability on timescales of $\sim 1\,$yr and no relativistic beaming, we find $T_B\approx 3000\,K$ at a distance of 1\,kpc (Galactic), or $T_B\approx 3\times 10^{15}\,$K at a distance of 1\,Gpc (extragalactic).  The former is consistent with a wide range of progenitors. In contrast, the latter exceeds the $10^{12}\,$K limit \citep{readhead94}, suggesting that if extragalactic the source is relativistic or the variability is not intrinsic.
 
 \subsubsection{Galactic source types}
At low frequencies, most potential Galactic transients emit via coherent processes that
 exhibit variability on
short timescales of seconds, minutes or hours: for example, giant pulses from pulsars, intermittent pulsars or flares from cool
stars or exoplanets \citep[e.g.,][]{bowman13}. For example, to have coherent emission with brightness temperature $>10^{12}\,$K would require emission on timescales $\lesssim 2000\,$s.
Given the timescales our search probes (15\,min for the TGSS observations, and a few hours for GLEAM), discovering a source with such short timescale variability would be very unlikely unless the duty cycle were quite large, and the unknown  timescale also means that any limits on brightness temperature are unconstraining.

\begin{figure}
\centering
\includegraphics[width=\columnwidth]{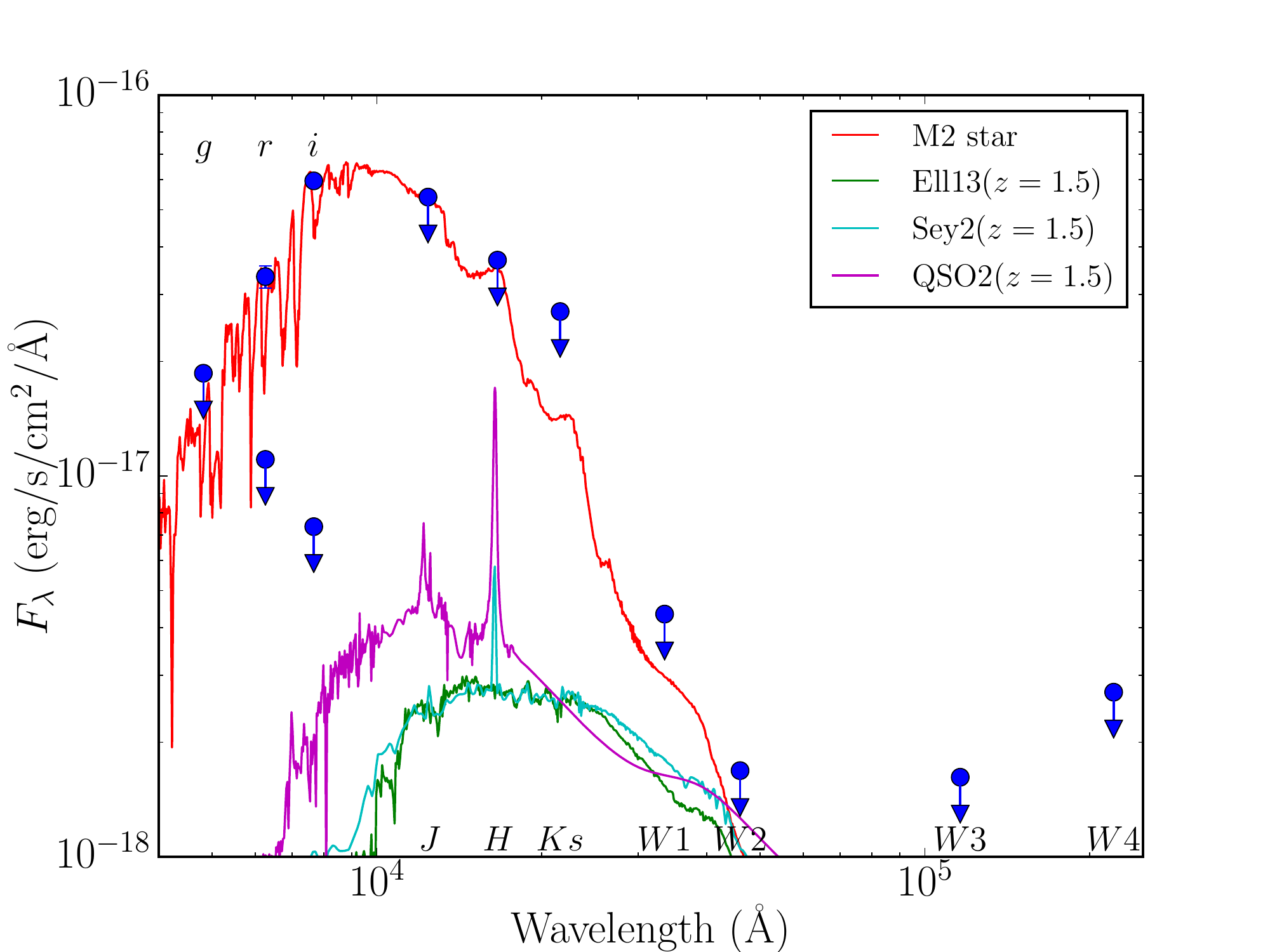}
\caption{Optical/infrared spectral energy distribution of \tgss, based on our SkyMapper $g$, $r$, and $i$ data long with archival 2MASS and \textit{WISE} upper limits (Tab.~\ref{t_oir}).  We show both the detected $r$ and $i$ measurements of the potential stellar source from Fig.~\ref{f_skymapper} as well as upper limits for the rest of the error region.  We  show a reddened M2 star from \citet{castelli04} at a distance of 1.1\,kpc (red curve), which passes through the $r$ and $i$ detections.  We also show an early-type galaxy (model ``Ell13"; green curve), a Seyfert galaxy (model ``Sey2"; cyan curve), and a QSO (model ``QSO2"; magenta curve) from \citet{polletta07}, all at redshift of 1.5 and normalized to $K_s=18\,$mag (following the $K-z$ relation of \citealt{willott03}).
}
\label{f_oir}
\end{figure}

If the optical source from \S~\ref{s_mw}  is not the counterpart of \tgss, we can exclude a range of  stellar \citep{covey07} or sub-stellar \citep{hawley02} counterparts.  Main-sequence stars with spectral types earlier than G8 can be excluded out to distances $\gtrsim20\,$kpc, and even stars as late as M6 can be excluded out to 1\,kpc.  Brown dwarfs can be excluded to distances of almost 1\,kpc (late M/early L) down to $\approx 10\,$pc (late L/early T).

If the source is the counterpart, the $r-i$ color implies a spectral type of roughly M2 at a distance of about 1\,kpc.  This is largely consistent with the 2MASS near-infrared upper limits (Fig.~\ref{f_oir}).  Such an association is plausible: while most low radio frequency observations of the lowest-mass stars and brown dwarfs have not seen any emission to much deeper limits \citep[e.g.,][]{jaeger11}, bright, short-duration flares have been seen from some higher mass M-dwarf stars \citep[e.g.,][]{spangler76}.  If a stellar flare, the implied luminosity density is about $4\times 10^{20}\,{\rm erg\,s}^{-1}\,{\rm Hz}^{-1}$, which is not out of line with observed flares \citep[e.g.,][]{nelson79}.
Yet given the flare rates of even the most active M dwarfs, and assuming the probability of burst emission follows Poisson statistics, the likelihood of observing a flare in the 15 minute TGSS observation is low ($<$20$\%$) \citep{abada-simon97}, and it is even more unlikely given the lack of flaring in the multiple epochs of MWA data we investigated.  In addition, this source is otherwise undistinguished, and we find a density of sources with $i<18.7$ of $0.0035\,{\rm arcsec}^{-2}$ implying $0.4$ sources within $6\arcsec$ by chance.    Searches for further flares from this source as well as more detailed investigations of the potential stellar counterpart are needed.

Another possibility is coherent emission from the magnetosphere of a Jupiter-like exoplanet \citep{zarka01}. The frequency of this type of emission is strongly tied to the magnetic field strength of the planet's magnetosphere: emission at 150\,MHz implies a magnetic field of about 50\,G.
Therefore, the observed radio emission could extend over a small bandwidth and  explain the non-detections at higher frequencies. However, recent modeling of the radio emission expected from Jupiter-like exoplanets predict flux densities of a few mJy \citep{fujii16, griessmeier11}, several orders of magnitude lower than the flux density measured for \tgss.

\subsubsection{Extragalactic source types}

In contrast to the Galactic sources, 
the sources expected to appear or disappear on the yearly timescales we probed in this survey are most likely to be extragalactic
synchrotron sources, such as afterglows from gamma-ray bursts or tidal disruption events, which tend to be 
relatively faint at low frequencies \citep{burlon15,metzger15}.  Note that synchrotron emission is also seen from Galactic X-ray binaries \citep[e.g.,][]{fender06}, but the lack of any high-energy counterpart makes this very unlikely.  

The best predictions for the rates of extragalactic synchrotron transients are given by \citet{metzger15}. For 
flux densities normalised to $z\approx0.55$, all the main categories of sources they consider (short gamma-ray 
bursts, long gamma-ray bursts, radio supernova, tidal disruption events and neutron star mergers) are expected to have flux densities of less than 0.1~mJy on timescales of 1 to 3 years. Note that we do not see any gamma-ray burst from either \textit{Swift} or \textit{Fermi} with a position consistent with \tgss.
If the object we have detected is an extragalactic synchrotron source
this could suggest a higher rate of these sources than predicted by \citet{metzger15}.

Another possibility is that the emission we see is related to an active galaxy, either intrinsic variability from an AGN flare \citep[e.g.,][]{hughes92,croft13,bignall15} or extrinsic variability caused by interstellar scintillation.  Given the large modulation implied by the GLEAM upper limit and the non-detections at other radio frequencies we can rule out refractive scintillation as the origin of the variability, 
but diffractive scintillation is possible. The NE2001 electron density model \citep{cordes02} predicts a scintillation bandwidth of $<0.1\,$kHz at 150\,MHz, suggesting that any scintillations are likely to be washed out.  
Scintillation by discrete structures like Extreme Scattering Events \citep[e.g.][]{fiedler87b} is still possible, but the degree of modulation needed to accomodate our 9\,GHz limits with a standard extragalactic SED (roughly 10) is significantly higher than is seen in most ESEs ($\sim 2$).

Assuming an extragalactic origin, we can use the photometry in Table~\ref{t_oir} to put limits on the host galaxy. Radio-loud AGN tend to be hosted by massive galaxies, and follow a rather tight relationship between $K$-magnitude and redshift \citep{willott03}. To be consistent with all of our data and upper limits requires models with $K \gtrsim 18$ (Fig.~\ref{f_oir}), which implies $z \gtrsim 1.5$ for typical (few $L_\star$) AGN hosts.
 The \textit{WISE} W1 photometry is suggestive of a similar lower bound in redshift \citep{gurkan14}, and the optical upper limits are also consistent with  a radio-loud AGN at moderate to high redshift \citep{stern12}. Overall we see a consistent  interpretation as  a $z \sim 1.5$ radio galaxy somewhat above  the break in the radio luminosity function \citep{vardoulaki10},
  although a less massive host could be at somewhat lower redshift.  Deeper near-IR imaging could help to identify any host galaxy, and deeper radio imaging could be used to search for a quiescent counterpart.

We plan to conduct further follow-up observations to establish whether
our candidate is an extragalactic synchrotron source. For example, we would expect
to detect the quiescent emission of a flaring AGN with deeper radio 
observations, or the host galaxy of a gamma-ray burst or supernovae with deeper
optical and infrared observations.

\subsection{Transient rates}
\begin{figure*}
\centering
\includegraphics[width=19cm]{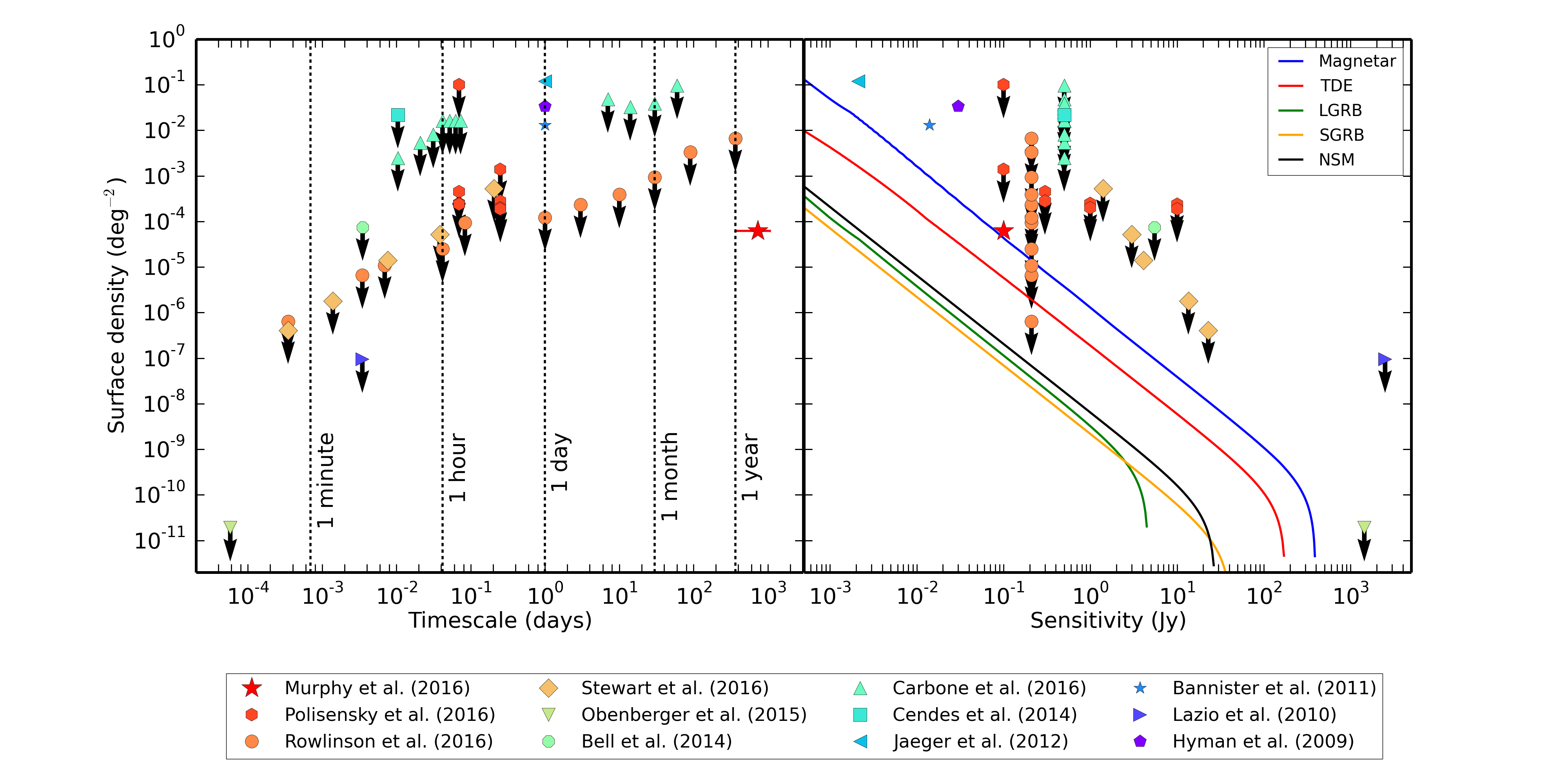}
\caption{Limits on the transient rates from our survey compared to previously published results at
low frequency ($<1$~GHz). The result presented in this paper is shown as a red star. The red error bar
shows the uncertainty in timescale, as our data spans 1 to 3 years. The coloured lines show the sky density of sources
above flux density $F_\nu$ for frequency $\nu = 150$~MHz. We have included predicted source rates for various source classes from Figure~3 in \citet{metzger15}, specifically: magnetars (blue); off-axis tidal disruption events (red); long GRBs with $\theta_{\rm obs} = 1.57$ (green); off-axis short GRBs (orange); and neutron star merger leaving black hole (black). See \citet{metzger15} for a detailed description of how these model
predictions were calculated.}\label{rates}
\end{figure*}

The total area covered by our search is 16\,230 sq deg. We detected a single 
transient above 100~mJy over this area, which results in a surface density 
estimate of $\rho = (6.2\pm6) \times10^{-5}$~deg$^{-2}$.
Figure~\ref{rates} shows our new result (red star) compared to other results 
from the literature.
The timescales explored by our survey range from 1 to 3 years (TGSS was 
conducted between 2011 April and 2012 March, and GLEAM was conducted between 
2013 June and 2014 July). To accommodate this uncertainty we have 
plotted our point at 2 years, with an error bar of $\pm1$~year.
\nocite{obenberger15}
\nocite{bell14}
\nocite{carbone16}
\nocite{cendes14}
\nocite{jaeger12}
\nocite{bannister11}
\nocite{lazio10}
\nocite{hyman09}

In terms of long timescales, the best limits in the literature are by \citet{rowlinson16} who found a 
surface density of $\rho < 6.6\times10^{-3}$~deg$^{-2}$ on yearly timescales. Our detection is 
an order 
of magnitude lower than this and so is consistent with previous results.
To date, most low-frequency surveys have been limited in sensitivity, something the expanded MWA, the JVLA and LOFAR will help overcome.

Figure~\ref{rates} also shows the predicted rates for a range of phenomena, as calculated by \citet{metzger15}. The coloured lines show the sky density of sources
above flux density $F_\nu$ for frequency $\nu = 150$~MHz. We have included a selection of results from Figure~3 in \citet{metzger15}, specifically: magnetars (blue); off-axis tidal disruption events (red); long GRBs with $\theta_{\rm obs} = 1.57$ (green); off-axis short GRBs (orange); and neutron star merger leaving black hole (black). See \citet{metzger15} for a detailed description of how these model
predictions were calculated, and their associated uncertainties, which can be up
to an order of magnitude. From plot we can see that our current results and
the limits set by \citet{rowlinson16} are now approaching the point of being able
to test these predictions.

\subsection{Search completeness}
It is possible that other transient sources were overlooked due to them being a single source at the resolution of 
TGSS but blended with another source at the resolution of GLEAM (in other words it is close, in sky projection, 
to a steady source). Alternatively, a transient source could be overlooked in our search if it occured at 
the same location as a persistent source: a recent example of this situation is the claimed host galaxy of a 
fast radio burst \citep{keane16} (but see \citealt{williams16} for further discussion).

This would have the effect of increasing or decreasing the flux density of 
a GLEAM source, but it would not be detected in the analysis we have done here. To formally account for 
this in our transient rate calculation we could exclude the area around each unresolved GLEAM source
in our survey region. This is equivalent to $A = 267\,860 \times 2.2^\prime \times 2.5^\prime = 409$~deg$^{-2}$, which is less than $3\%$ of the survey area.

Another possibility is that some of the GLEAM sources with a TGSS match were false matches due to positional coincidence
of the source in the TGSS catalogue. To evaluate this we shifted the positions of the GLEAM sources by a random offset
in right ascension and declination of between 5 and 10 arc minutes. We then repeated the cross-matching with the TGSS
catalogue and found that of the 99\,658 compact sources above 100~mJy in our survey region, 1371 had a match. 
This implies that up to $\sim 1.3\%$ of the sources ruled out in our process could have been transients. 
However, since the total number of transients in our survey is $\ll 1$, this means that the expectation is that
the number of sources we would miss due to this issue is less than one. The caveat here is that many classes 
of physical transient (e.g. radio supernovae) will occur at the same location as a persistent radio source 
(the host galaxy) and this is not accounted for in this analysis.

\subsection{Catalogue reliability}
The small number of candidate transient sources detected in our search reflects the high reliability
of both the GLEAM and TGSS catalogues for point sources with flux densities above $100-200$~mJy. The two GLEAM 
candidates 
GLEAM~J153424$-$114947 and GLEAM~J153653$-$115052 have been removed from the publicly available 
GLEAM catalogue. We note that this introduces a small bias in the GLEAM catalogue, since there will be
similar artefacts that are weaker than our flux density cutoff that have not been removed.

The importance of rigorous analysis of 
imaging data to confirm or rule out candidates was demonstrated by \citet{frail12} who re-analysed 
VLA archival data to rule out the long-standing \citet{bower07} candidates.
Our work provides a good example of this; in an attempt to confirm our transient candidates we 
reprocessed the relevant data for both GLEAM and TGSS, and in two cases established that the 
candidates were imaging artefacts.

\section{Conclusions}\label{s_conclusion}
We have conducted a blind search for low frequency radio transients by comparing the GLEAM and TGSS~ADR1
surveys at 200 and 147.5~MHz respectively. 
From a total sample of 234\,333 GLEAM sources and 275\,612 TGSS ADR1 sources in the overlap region 
between the two surveys, there were 99\,658 GLEAM sources and 38\,978 TGSS ADR sources that passed our 
flux density cutoff and compactness criteria.
From this sub-sample we found three candidate transient sources, but further analysis 
identified two of these as imaging artefacts. 
We present one candidate transient: \tgss, that has a flux density of 304~mJy at 147.5~MHz (scaled to $182\pm26$~mJy 
based on a flux scale comparison with GLEAM in the region of the source). This source was not detected in the GLEAM
survey 3 years later, implying it had faded to below $<41$~mJy. 
It was also not detected in other archival radio data, or in our ATCA 
observations. 

Based on this detection we calculated a surface density 
estimate of $\rho = 6.2\times10^{-5}$~deg$^{-2}$ for low-frequency radio transients
on timescales of 1 to 3 years.

It is worth noting that the distinction between transients and extremely variable sources can be merely
a selection effect due to the limited sensitivity and sampling of a given survey. A more comprehensive 
analysis of variability in this dataset is ongoing. We are also conducting 
further follow-up observations of \tgss\ in order to classify it.

\section*{Acknowledgments}
We thank David McConnell and Richard Hunstead
for providing archival radio data from the ATPMN and Molonglo 408~MHz surveys, respectively. We also thank Brian Metzger for providing the
model predictions for Figure 8.
This research was conducted by the Australian Research Council Centre of Excellence for All-sky Astrophysics (CAASTRO), 
through project number CE110001020.
DLK and SDC are additionally supported by NSF grant AST-1412421.
This research work has used the TIFR GMRT Sky Survey (http://tgss.ncra.tifr.res.in) data products. We thank the staff of the GMRT that made these observations possible. GMRT is run by the National Centre for Radio Astrophysics of the Tata Institute of Fundamental Research.
The national facility capability for SkyMapper has been funded through ARC LIEF grant LE130100104 from the Australian Research Council, awarded to the University of Sydney, the Australian National University, Swinburne University of Technology, the University of Queensland, the University of Western Australia, the University of Melbourne, Curtin University of Technology, Monash University and the Australian Astronomical Observatory. SkyMapper is owned and operated by The Australian National University's Research School of Astronomy and Astrophysics.

This scientific work makes use of the Murchison Radio-astronomy Observatory, operated by CSIRO. We acknowledge the Wajarri Yamatji people as the traditional owners of the Observatory site. Support for the operation of the MWA is provided by the Australian Government (NCRIS), under a contract to Curtin University administered by Astronomy Australia Limited. We acknowledge the Pawsey Supercomputing Centre which is supported by the Western Australian and Australian Governments.

\bibliographystyle{mn2e}
\bibliography{mn-jour,mwa,pulsars,vast}

\begin{thebibliography}{}

\bibitem[\protect\citeauthoryear{{Abada-Simon} \& {Aubier}}{{Abada-Simon} \&
  {Aubier}}{1997}]{abada-simon97}
{Abada-Simon} M.,  {Aubier} M.,  1997, A\&AS, 125

\bibitem[\protect\citeauthoryear{{Baars}, {Genzel}, {Pauliny-Toth} \&
  {Witzel}}{{Baars} et~al.}{1977}]{baars77}
{Baars} J.~W.~M.,  {Genzel} R.,  {Pauliny-Toth} I.~I.~K.,    {Witzel} A.,
  1977, A\&A, 61, 99

\bibitem[\protect\citeauthoryear{{Bannister}, {Murphy}, {Gaensler}, {Hunstead}
  \& {Chatterjee}}{{Bannister} et~al.}{2011}]{bannister11}
{Bannister} K.~W.,  {Murphy} T.,  {Gaensler} B.~M.,  {Hunstead} R.~W.,
  {Chatterjee} S.,  2011, MNRAS, 412, 634

\bibitem[\protect\citeauthoryear{{Bell} et~al.,}{{Bell}  et~al.}{2014}]{bell14}
{Bell} M.~E.,  et~al., 2014, MNRAS, 438, 352

\bibitem[\protect\citeauthoryear{{Berger}}{{Berger}}{2006}]{berger06}
{Berger} E.,  2006, ApJ, 648, 629

\bibitem[\protect\citeauthoryear{{Bignall}, {Croft}, {Hovatta}, {Koay},
  {Lazio}, {Macquart} \& {Reynolds}}{{Bignall} et~al.}{2015}]{bignall15}
{Bignall} H.~E.,  {Croft} S.,  {Hovatta} T.,  {Koay} J.~Y.,  {Lazio} J.,
  {Macquart} J.~P.,    {Reynolds} C.,  2015, Advancing Astrophysics with the
  Square Kilometre Array (AASKA14), p.~58

\bibitem[\protect\citeauthoryear{{Bower} et~al.,}{{Bower}
  et~al.}{2007}]{bower07}
{Bower} G.~C.,  et~al., 2007, ApJ, 666, 346

\bibitem[\protect\citeauthoryear{{Bowman} et~al.,}{{Bowman}
  et~al.}{2013}]{bowman13}
{Bowman} J.~D.,  et~al., 2013, PASA, 30, 31

\bibitem[\protect\citeauthoryear{{Burlon}, {Ghirlanda}, {van der Horst},
  {Murphy}, {Wijers}, {Gaensler}, {Ghisellini} \& {Prandoni}}{{Burlon}
  et~al.}{2015}]{burlon15}
{Burlon} D.,  {Ghirlanda} G.,  {van der Horst} A.,  {Murphy} T.,  {Wijers}
  R.~A.~M.~J.,  {Gaensler} B.,  {Ghisellini} G.,    {Prandoni} I.,  2015,
  Advancing Astrophysics with the Square Kilometre Array (AASKA14), p.~52

\bibitem[\protect\citeauthoryear{{Carbone} et~al.,}{{Carbone}
  et~al.}{2016}]{carbone16}
{Carbone} D.,  et~al., 2016, MNRAS, 459, 3161

\bibitem[\protect\citeauthoryear{{Castelli} \& {Kurucz}}{{Castelli} \&
  {Kurucz}}{2004}]{castelli04}
{Castelli} F.,  {Kurucz} R.~L.,  2004, in Piskunov N.,  et~al. eds, {IAU Symp.
  No 210, Modelling of Stellar Atmospheres} {New Grids of ATLAS9 Model
  Atmospheres}

\bibitem[\protect\citeauthoryear{{Cendes} et~al.,}{{Cendes}
  et~al.}{2014}]{cendes14}
{Cendes} Y.,  et~al., 2014, arXiv:1412.3986

\bibitem[\protect\citeauthoryear{{Clarke}, {Kassim}, {Polisensky}, {Peters},
  {Giacintucci} \& {Hyman}}{{Clarke} et~al.}{2016}]{clarke16}
{Clarke} T.,  {Kassim} N.,  {Polisensky} E.,  {Peters} W.,  {Giacintucci} S.,
   {Hyman} S.~D.,  2016, arXiv:1603.03080

\bibitem[\protect\citeauthoryear{{Condon}, {Cotton}, {Greisen}, {Yin},
  {Perley}, {Taylor} \& {Broderick}}{{Condon} et~al.}{1998}]{condon98}
{Condon} J.~J.,  {Cotton} W.~D.,  {Greisen} E.~W.,  {Yin} Q.~F.,  {Perley}
  R.~A.,  {Taylor} G.~B.,    {Broderick} J.~J.,  1998, AJ, 115, 1693

\bibitem[\protect\citeauthoryear{{Cordes} et~al.,}{{Cordes}
  et~al.}{2004}]{cordes04}
{Cordes} J.~M.,  et~al., 2004, New Astronomy Reviews, 48, 1459

\bibitem[\protect\citeauthoryear{{Cordes} \& {Lazio}}{{Cordes} \&
  {Lazio}}{2002}]{cordes02}
{Cordes} J.~M.,  {Lazio} T.~J.~W.,  2002, arXiv:0207156

\bibitem[\protect\citeauthoryear{{Covey}, {Ivezi{\'c}}, {Schlegel},
  {Finkbeiner}, {Padmanabhan}, {Lupton}, {Ag{\"u}eros}, {Bochanski}, {Hawley},
  {West}, {Seth}, {Kimball}, {Gogarten}, {Claire}, {Haggard}, {Kaib},
  {Schneider} \& {Sesar}}{{Covey} et~al.}{2007}]{covey07}
{Covey} K.~R.,  {Ivezi{\'c}} {\v Z}.,  {Schlegel} D.,  {Finkbeiner} D.,
  {Padmanabhan} N.,  {Lupton} R.~H.,  {Ag{\"u}eros} M.~A.,  {Bochanski} J.~J.,
  {Hawley} S.~L.,  {West} A.~A.,  {Seth} A.,  {Kimball} A.,  {Gogarten} S.~M.,
  {Claire} M.,  {Haggard} D.,  {Kaib} N.,  {Schneider} D.~P.,    {Sesar} B.,
  2007, AJ, 134, 2398

\bibitem[\protect\citeauthoryear{{Croft}, {Bower} \& {Whysong}}{{Croft}
  et~al.}{2013}]{croft13}
{Croft} S.,  {Bower} G.~C.,    {Whysong} D.,  2013, ApJ, 762, 93

\bibitem[\protect\citeauthoryear{{Drimmel}, {Cabrera-Lavers} \&
  {L{\'o}pez-Corredoira}}{{Drimmel} et~al.}{2003}]{drimmel03}
{Drimmel} R.,  {Cabrera-Lavers} A.,    {L{\'o}pez-Corredoira} M.,  2003, A\&A,
  409, 205

\bibitem[\protect\citeauthoryear{{Fender}}{{Fender}}{2006}]{fender06}
{Fender} R.,  2006, {Jets from X-ray binaries}.
pp 381--419

\bibitem[\protect\citeauthoryear{{Fender}, {Stewart}, {Macquart}, {Donnarumma},
  {Murphy}, {Deller}, {Paragi} \& {Chatterjee}}{{Fender}
  et~al.}{2015}]{fender15}
{Fender} R.,  {Stewart} A.,  {Macquart} J.-P.,  {Donnarumma} I.,  {Murphy} T.,
  {Deller} A.,  {Paragi} Z.,    {Chatterjee} S.,  2015, arXiv:1507.00729

\bibitem[\protect\citeauthoryear{{Fiedler} et~al.,}{{Fiedler}
  et~al.}{1987}]{fiedler87b}
{Fiedler} R.~L.,  et~al., 1987, Nature, 326, 675

\bibitem[\protect\citeauthoryear{{Frail}, {Kulkarni}, {Ofek}, {Bower} \&
  {Nakar}}{{Frail} et~al.}{2012}]{frail12}
{Frail} D.~A.,  {Kulkarni} S.~R.,  {Ofek} E.~O.,  {Bower} G.~C.,    {Nakar} E.,
   2012, ApJ, 747, 70

\bibitem[\protect\citeauthoryear{{Fujii}, {Spiegel}, {Mroczkowski}, {Nordhaus},
  {Zimmerman}, {Parsons}, {Mirbabayi} \& {Madhusudhan}}{{Fujii}
  et~al.}{2016}]{fujii16}
{Fujii} Y.,  {Spiegel} D.~S.,  {Mroczkowski} T.,  {Nordhaus} J.,  {Zimmerman}
  N.~T.,  {Parsons} A.~R.,  {Mirbabayi} M.,    {Madhusudhan} N.,  2016, ApJ,
  820, 122

\bibitem[\protect\citeauthoryear{{Grie{\ss}meier}, {Zarka} \&
  {Girard}}{{Grie{\ss}meier} et~al.}{2011}]{griessmeier11}
{Grie{\ss}meier} J.-M.,  {Zarka} P.,    {Girard} J.~N.,  2011, Radio Science,
  46, 0

\bibitem[\protect\citeauthoryear{{G{\"u}rkan}, {Hardcastle} \&
  {Jarvis}}{{G{\"u}rkan} et~al.}{2014}]{gurkan14}
{G{\"u}rkan} G.,  {Hardcastle} M.~J.,    {Jarvis} M.~J.,  2014, MNRAS, 438,
  1149

\bibitem[\protect\citeauthoryear{{Hancock}, {Gaensler} \& {Murphy}}{{Hancock}
  et~al.}{2011}]{hancock11}
{Hancock} P.,  {Gaensler} B.~M.,    {Murphy} T.,  2011, ApJ, 735, L35

\bibitem[\protect\citeauthoryear{{Hawley} et~al.,}{{Hawley}
  et~al.}{2002}]{hawley02}
{Hawley} S.~L.,  et~al., 2002, AJ, 123, 3409

\bibitem[\protect\citeauthoryear{{Hess} \& {Zarka}}{{Hess} \&
  {Zarka}}{2011}]{hess11}
{Hess} S.~L.~G.,  {Zarka} P.,  2011, A\&A, 531, A29

\bibitem[\protect\citeauthoryear{{Hughes}, {Aller} \& {Aller}}{{Hughes}
  et~al.}{1992}]{hughes92}
{Hughes} P.~A.,  {Aller} H.~D.,    {Aller} M.~F.,  1992, ApJ, 396, 469

\bibitem[\protect\citeauthoryear{{Hurley-Walker} et~al.,}{{Hurley-Walker}
  et~al.}{2016}]{hurleywalker16}
{Hurley-Walker} N.,  et~al., 2016, Submitted

\bibitem[\protect\citeauthoryear{{Hyman} et~al.,}{{Hyman}
  et~al.}{2005}]{hyman05}
{Hyman} S.~D.,  et~al., 2005, Nat, 434, 50

\bibitem[\protect\citeauthoryear{{Hyman}, {Wijnands}, {Lazio}, {Pal},
  {Starling}, {Kassim} \& {Ray}}{{Hyman} et~al.}{2009}]{hyman09}
{Hyman} S.~D.,  {Wijnands} R.,  {Lazio} T.~J.~W.,  {Pal} S.,  {Starling} R.,
  {Kassim} N.~E.,    {Ray} P.~S.,  2009, ApJ, 696, 280

\bibitem[\protect\citeauthoryear{{Intema} et~al.,}{{Intema}
  et~al.}{2009}]{intema09}
{Intema} H.~T.,  et~al., 2009, A\&A, 501, 1185

\bibitem[\protect\citeauthoryear{{Intema}, {Jagannathan}, {Mooley} \&
  {Frail}}{{Intema} et~al.}{2016}]{intema16}
{Intema} H.~T.,  {Jagannathan} P.,  {Mooley} K.~P.,    {Frail} D.~A.,  2016,
  arXiv:1603.04368

\bibitem[\protect\citeauthoryear{{Jaeger}, {Hyman}, {Kassim} \&
  {Lazio}}{{Jaeger} et~al.}{2012}]{jaeger12}
{Jaeger} T.~R.,  {Hyman} S.~D.,  {Kassim} N.~E.,    {Lazio} T.~J.~W.,  2012,
  AJ, 143, 96

\bibitem[\protect\citeauthoryear{{Jaeger}, {Osten}, {Lazio}, {Kassim} \&
  {Mutel}}{{Jaeger} et~al.}{2011}]{jaeger11}
{Jaeger} T.~R.,  {Osten} R.~A.,  {Lazio} T.~J.,  {Kassim} N.,    {Mutel} R.~L.,
   2011, AJ, 142, 189

\bibitem[\protect\citeauthoryear{{Kaplan} et~al.,}{{Kaplan}
  et~al.}{2015}]{kaplan15}
{Kaplan} D.~L.,  et~al., 2015, ApJ, 809, L12

\bibitem[\protect\citeauthoryear{{Keane} et~al.,}{{Keane}
  et~al.}{2016}]{keane16}
{Keane} E.~F.,  et~al., 2016, Nat, 530, 453

\bibitem[\protect\citeauthoryear{{Keller} et~al.,}{{Keller}
  et~al.}{2007}]{keller07}
{Keller} S.~C.,  et~al., 2007, PASA, 24, 1

\bibitem[\protect\citeauthoryear{{Lane}, {Cotton}, {van Velzen}, {Clarke},
  {Kassim}, {Helmboldt}, {Lazio} \& {Cohen}}{{Lane} et~al.}{2014}]{lane14}
{Lane} W.~M.,  {Cotton} W.~D.,  {van Velzen} S.,  {Clarke} T.~E.,  {Kassim}
  N.~E.,  {Helmboldt} J.~F.,  {Lazio} T.~J.~W.,    {Cohen} A.~S.,  2014, MNRAS,
  440, 327

\bibitem[\protect\citeauthoryear{{Lazio}, {Shankland}, {Farrell} \&
  {Blank}}{{Lazio} et~al.}{2010}]{lazio10}
{Lazio} T.~J.~W.,  {Shankland} P.~D.,  {Farrell} W.~M.,    {Blank} D.~L.,
  2010, AJ, 140, 1929

\bibitem[\protect\citeauthoryear{{Loi} et~al.,}{{Loi}  et~al.}{2015}]{loi15b}
{Loi} S.~T.,  et~al., 2015, MNRAS, 453, 2731

\bibitem[\protect\citeauthoryear{{Mauch}, {Murphy}, {Buttery}, {Curran},
  {Hunstead}, {Piestrzynski}, {Robertson} \& {Sadler}}{{Mauch}
  et~al.}{2003}]{mauch03}
{Mauch} T.,  {Murphy} T.,  {Buttery} H.~J.,  {Curran} J.,  {Hunstead} R.~W.,
  {Piestrzynski} B.,  {Robertson} J.~G.,    {Sadler} E.~M.,  2003, MNRAS, 342,
  1117

\bibitem[\protect\citeauthoryear{{McConnell}, {Sadler}, {Murphy} \&
  {Ekers}}{{McConnell} et~al.}{2012}]{mcconnell12}
{McConnell} D.,  {Sadler} E.~M.,  {Murphy} T.,    {Ekers} R.~D.,  2012, MNRAS,
  422, 1527

\bibitem[\protect\citeauthoryear{{Metzger}, {Williams} \& {Berger}}{{Metzger}
  et~al.}{2015}]{metzger15}
{Metzger} B.~D.,  {Williams} P.~K.~G.,    {Berger} E.,  2015, ApJ, 806, 224

\bibitem[\protect\citeauthoryear{{Murphy} et~al.,}{{Murphy}
  et~al.}{2015}]{murphy15}
{Murphy} T.,  et~al., 2015, MNRAS, 446, 2560

\bibitem[\protect\citeauthoryear{{Nelson}, {Robinson}, {Slee}, {Fielding},
  {Page} \& {Walker}}{{Nelson} et~al.}{1979}]{nelson79}
{Nelson} G.~J.,  {Robinson} R.~D.,  {Slee} O.~B.,  {Fielding} G.,  {Page}
  A.~A.,    {Walker} W.~S.~G.,  1979, MNRAS, 187, 405

\bibitem[\protect\citeauthoryear{{Obenberger} et~al.,}{{Obenberger}
  et~al.}{2015}]{obenberger15}
{Obenberger} K.~S.,  et~al., 2015, Journal of Astronomical Instrumentation, 4,
  1550004

\bibitem[\protect\citeauthoryear{{Polisensky}, {Lane}, {Hyman}, {Kassim},
  {Giacintucci}, {Clarke}, {Cotton}, {Cleland} \& {Frail}}{{Polisensky}
  et~al.}{2016}]{polisensky16}
{Polisensky} E.,  {Lane} W.~M.,  {Hyman} S.~D.,  {Kassim} N.~E.,  {Giacintucci}
  S.,  {Clarke} T.~E.,  {Cotton} W.~D.,  {Cleland} E.,    {Frail} D.~A.,  2016,
  arXiv:1604.00667

\bibitem[\protect\citeauthoryear{{Polletta} et~al.,}{{Polletta}
  et~al.}{2007}]{polletta07}
{Polletta} M.,  et~al., 2007, ApJ, 663, 81

\bibitem[\protect\citeauthoryear{{Readhead}}{{Readhead}}{1994}]{readhead94}
{Readhead} A.~C.~S.,  1994, ApJ, 426, 51

\bibitem[\protect\citeauthoryear{{Rowlinson} et~al.,}{{Rowlinson}
  et~al.}{2016}]{rowlinson16}
{Rowlinson} A.,  et~al., 2016, MNRAS, 458, 3506

\bibitem[\protect\citeauthoryear{{Scaife} \& {Heald}}{{Scaife} \&
  {Heald}}{2012}]{scaife12}
{Scaife} A.~M.~M.,  {Heald} G.~H.,  2012, MNRAS, 423, L30

\bibitem[\protect\citeauthoryear{{Skrutskie} et~al.,}{{Skrutskie}
  et~al.}{2006}]{twomass}
{Skrutskie} M.~F.,  et~al., 2006, AJ, 131, 1163

\bibitem[\protect\citeauthoryear{{Sobey} et~al.,}{{Sobey}
  et~al.}{2015}]{sobey15}
{Sobey} C.,  et~al., 2015, MNRAS, 451, 2493

\bibitem[\protect\citeauthoryear{{Spangler} \& {Moffett}}{{Spangler} \&
  {Moffett}}{1976}]{spangler76}
{Spangler} S.~R.,  {Moffett} T.~J.,  1976, ApJ, 203, 497

\bibitem[\protect\citeauthoryear{{Stern}, {Assef}, {Benford}, {Blain}, {Cutri},
  {Dey}, {Eisenhardt}, {Griffith}, {Jarrett}, {Lake}, {Masci}, {Petty},
  {Stanford}, {Tsai}, {Wright}, {Yan}, {Harrison} \& {Madsen}}{{Stern}
  et~al.}{2012}]{stern12}
{Stern} D.,  {Assef} R.~J.,  {Benford} D.~J.,  {Blain} A.,  {Cutri} R.,  {Dey}
  A.,  {Eisenhardt} P.,  {Griffith} R.~L.,  {Jarrett} T.~H.,  {Lake} S.,
  {Masci} F.,  {Petty} S.,  {Stanford} S.~A.,  {Tsai} C.-W.,  {Wright} E.~L.,
  {Yan} L.,  {Harrison} F.,    {Madsen} K.,  2012, ApJ, 753, 30

\bibitem[\protect\citeauthoryear{{Stewart} et~al.,}{{Stewart}
  et~al.}{2016}]{stewart16}
{Stewart} A.~J.,  et~al., 2016, MNRAS, 456, 2321

\bibitem[\protect\citeauthoryear{{Swarup}}{{Swarup}}{1990}]{swarup90}
{Swarup} G.,  1990, Indian Journal of Radio and Space Physics, 19, 493

\bibitem[\protect\citeauthoryear{{Taylor} et~al.,}{{Taylor}
  et~al.}{2012}]{taylor12}
{Taylor} G.~B.,  et~al., 2012, Journal of Astronomical Instrumentation, 1,
  1250004

\bibitem[\protect\citeauthoryear{{Tingay} et~al.,}{{Tingay}
  et~al.}{2013}]{tingay13}
{Tingay} S.~J.,  et~al., 2013, PASA, 30, 7

\bibitem[\protect\citeauthoryear{{van Haarlem} et~al.,}{{van Haarlem}
  et~al.}{2013}]{vanhaarlem13}
{van Haarlem} M.~P.,  et~al., 2013, A\&A, 556, A2

\bibitem[\protect\citeauthoryear{{van Weeren} et~al.,}{{van Weeren}
  et~al.}{2016}]{vanweeren16}
{van Weeren} R.~J.,  et~al., 2016, ApJS, 223, 2

\bibitem[\protect\citeauthoryear{{Vardoulaki}, {Rawlings}, {Hill}, {Mauch},
  {Inskip}, {Riley}, {Brand}, {Croft} \& {Willott}}{{Vardoulaki}
  et~al.}{2010}]{vardoulaki10}
{Vardoulaki} E.,  {Rawlings} S.,  {Hill} G.~J.,  {Mauch} T.,  {Inskip} K.~J.,
  {Riley} J.,  {Brand} K.,  {Croft} S.,    {Willott} C.~J.,  2010, MNRAS, 401,
  1709

\bibitem[\protect\citeauthoryear{{Wayth} et~al.,}{{Wayth}
  et~al.}{2015}]{wayth15}
{Wayth} R.~B.,  et~al., 2015, PASA, 32, 25

\bibitem[\protect\citeauthoryear{{Williams} \& {Berger}}{{Williams} \&
  {Berger}}{2016}]{williams16}
{Williams} P.~K.~G.,  {Berger} E.,  2016, ApJ, 821, L22

\bibitem[\protect\citeauthoryear{{Willott}, {Rawlings}, {Jarvis} \&
  {Blundell}}{{Willott} et~al.}{2003}]{willott03}
{Willott} C.~J.,  {Rawlings} S.,  {Jarvis} M.~J.,    {Blundell} K.~M.,  2003,
  MNRAS, 339, 173

\bibitem[\protect\citeauthoryear{{Wright} et~al.,}{{Wright}
  et~al.}{2010}]{wright10}
{Wright} E.~L.,  et~al., 2010, AJ, 140, 1868

\bibitem[\protect\citeauthoryear{{Zarka}, {Treumann}, {Ryabov} \&
  {Ryabov}}{{Zarka} et~al.}{2001}]{zarka01}
{Zarka} P.,  {Treumann} R.~A.,  {Ryabov} B.~P.,    {Ryabov} V.~B.,  2001,
  Astrophys. Space. Sci., 277, 293

\end{thebibliography}

\label{lastpage}

\end{document}